\documentclass[conference]{IEEEtran}
\IEEEoverridecommandlockouts

\usepackage{cite}
\usepackage{amsmath,amssymb,amsfonts}
\usepackage{graphicx}
\usepackage{xcolor}
\usepackage{array,multirow}
\usepackage{tablefootnote}
\usepackage{threeparttable}
\usepackage{booktabs, makecell, tabularx}

\def\BibTeX{{\rm B\kern-.05em{\sc i\kern-.025em b}\kern-.08em
    T\kern-.1667em\lower.7ex\hbox{E}\kern-.125emX}}

\makeatletter
\IEEEtriggercmd{\reset@font\normalfont\fontsize{7.9pt}{8.60pt}\selectfont}
\newcommand*\titleheader[1]{\gdef\@titleheader{#1}}
\AtBeginDocument{%
  \let\st@red@title\@title
  \def\@title{%
    \bgroup\normalfont\normalsize\centering\@titleheader\par\egroup
    \vskip1ex\st@red@title}
}
\makeatother
\IEEEtriggeratref{1}

\titleheader{\vspace{-30pt}To appear at the 31st IEEE Int'l. Conference on Electronics Circuits and Systems (ICECS), 18-20 Nov, 2024, Nancy, France.}

\title{Development of High-Performance DSP Algorithms on the European Rad-Hard NG-ULTRA SoC FPGA%
\thanks{\hspace{-8pt}Work partially supported by the 
European Space Agency (ESA) via the activity ``QUEENS3'' with reference number 4000134874/21/NL/AR/vav.}}%

\makeatletter
\def\ps@IEEEtitlepagestyle{
  \def\@oddfoot{\mycopyrightnotice}
  \def\@evenfoot{}
}
\def\mycopyrightnotice{
  {\footnotesize
  \begin{minipage}{\textwidth}
  \centering%
  ~\copyright~2024 IEEE.  Personal use of this material is permitted.  Permission from IEEE must be obtained for all other uses, in any current or future media, including reprinting/republishing this material for advertising or promotional purposes, creating new collective works, for resale or redistribution\\to servers or lists, or reuse of any copyrighted component of this work in other works.
  \end{minipage}
  }
}
    
\begin{document}

\author{\IEEEauthorblockN{%
Vasileios Leon\IEEEauthorrefmark{1},
Anastasios Xynos\IEEEauthorrefmark{1},
Dimitrios Soudris\IEEEauthorrefmark{1}, 
George Lentaris\IEEEauthorrefmark{2}\IEEEauthorrefmark{1},\\[2pt] 
Ruben Domingo\IEEEauthorrefmark{3},
Arturo Perez\IEEEauthorrefmark{3},
David Gonzalez--Arjona\IEEEauthorrefmark{3}, 
Isabelle Conway\IEEEauthorrefmark{4},
David Merodio Codinachs\IEEEauthorrefmark{4}}\\[-9pt] 
\IEEEauthorblockA{\IEEEauthorrefmark{1}\emph{National Technical University of Athens, School of Electrical and Computer Engineering, Athens, Greece}}\\[-12pt] 
\IEEEauthorblockA{\IEEEauthorrefmark{2}\emph{University of West Attica, Department of Informatics and Computer Engineering, Athens, Greece}}\\[-12pt] 
\IEEEauthorblockA{\IEEEauthorrefmark{3}\emph{GMV Aerospace and Defence SAU, Flight Segment and Robotics Business Unit, Madrid, Spain}}\\[-12pt] 
\IEEEauthorblockA{\IEEEauthorrefmark{4}\emph{European Space Agency, European Space Research and Technology Centre, Noordwijk, Netherlands}}\\[-12pt] 
}

\maketitle

\begin{abstract}
The emergence of demanding space applications has modified the traditional landscape of computing systems in space. When reliability is a first-class concern, in addition to enhanced performance-per-Watt, radiation-hardened FPGAs are favored. In this context, the current paper evaluates the first European radiation-hardened SoC FPGA, i.e., NanoXplore's NG-ULTRA, for accelerating high-performance DSP algorithms from space applications. The proposed development \& testing methodologies provide efficient implementations, while they also aim to test the new NG-ULTRA hardware and its associated software tools. The results show that NG-ULTRA achieves competitive resource utilization and performance, constituting it as a very promising device for space missions, especially for Europe.   
\end{abstract}


\section{Introduction}
The space industry is continuously evaluating high-performance embedded platforms for handling the large data volumes and increased computational demands of novel space applications. 
The enhanced performance-per-Watt of FPGAs has constituted them as an attractive solution for efficient on-board payload data processing \cite{aiaa}, e.g., in applications for Vision-Based Navigation (VBN), Earth Observation (EO), and Satellite Communications (SatCom).
When the radiation, thermal \& vibration resilience
is of utmost importance, radiation-hardened/space-grade FPGAs are used to increase the reliability,
outperforming at the same time traditional processors (e.g., the CPU-based RAD750 and AT697F) in terms of processing speed. 

The market of high-density space-grade FPGAs 
is mainly occupied by two USA-based vendors; AMD/Xilinx (with Virtex-4QV, Virtex-5QV, and RT Kintex US) and Microchip/Microsemi (with RTAX, RTG4, and RT PolarFire).
It is also worth mentioning that there are only two space-grade SoC FPGAs available, 
i.e., AMD's XQR Versal and Microchip's RT PolarFire SoC.
Very recently, the European company NanoXplore entered this limited market by 
providing 
radiation-hardened FPGAs and SoC FPGAs \cite{nxx},
which are called BRAVE -- Big Re-programmable Arrays for Versatile Environments.  
These devices have been already 
considered for current and future space missions and projects \cite{spacebrave, ngmoon}.
NG-ULTRA \cite{ultra} belongs in the BRAVE FPGA family, and it is 
the first European 28-nm radiation-hardened SoC FPGA. 
It integrates programmable logic resources (500K LUT4, 32K RAM bits, 1.3K DSP)
and the DAHLIA SoC \cite{dahlia} featured with a
quad-core ARM-R52@600MHz processor. 

Over the last years, NanoXplore's radiation-hardened FPGAs 
have attracted the interest of industry and academia \cite{ahs_brave, access_brave, shyloc_fpga, zenoleon, klemen_new, cnn_brave, obc_ultra}.
The European Space Agency (ESA) is also carrying out a set of activities towards testing, evaluating, and improving the European space-grade FPGAs. 
In this context, 
we devise a methodology to 
support and test the development on these FPGAs,
considering that they are newly released, 
their hardware is radiation-hardened
and their software tools are still evolving. 
The proposed development \& testing methodology is based on systematic tool- and HDL-level exploration, aiming to deliver high-performance, resource-efficient implementations for NG-ULTRA. 
The implemented hardware kernels are compute- and memory-intensive DSP algorithms from the computer vision and signal processing domains, which are employed in realistic  space applications on-board rovers/spacecraft. 

The current paper contributes as follows:
(i) it introduces an enhanced version of our methodology,
evolved from our previous works \cite{ahs_brave, access_brave}, 
for development \& testing on new space-grade FPGAs,
(ii) it evaluates demanding, space-representative implementations on NG-ULTRA compared to well-established FPGAs,
and (iii) it reports benchmarking results for NG-ULTRA's processing system. 
According to our experimental evaluation, 
the resource utilization on NG-ULTRA is comparable to that of the competitors,
while the performance is sufficient (also given that NG-ULTRA is radiation-hardened-by-design).
Finally, as shown, the results for the processing system fit within the expected margin. 

\section{Development \& Testing Methodology}

The proposed methodology is built around 
NanoXplore's programmable-logic 
toolflow (synthesis and place \& route with the Impulse SW tool),
as well as it performs benchmarking of the processing system.

\subsection{SoC's Programmable Logic}

\begin{figure}[t]
\centering
\includegraphics[width=\columnwidth]{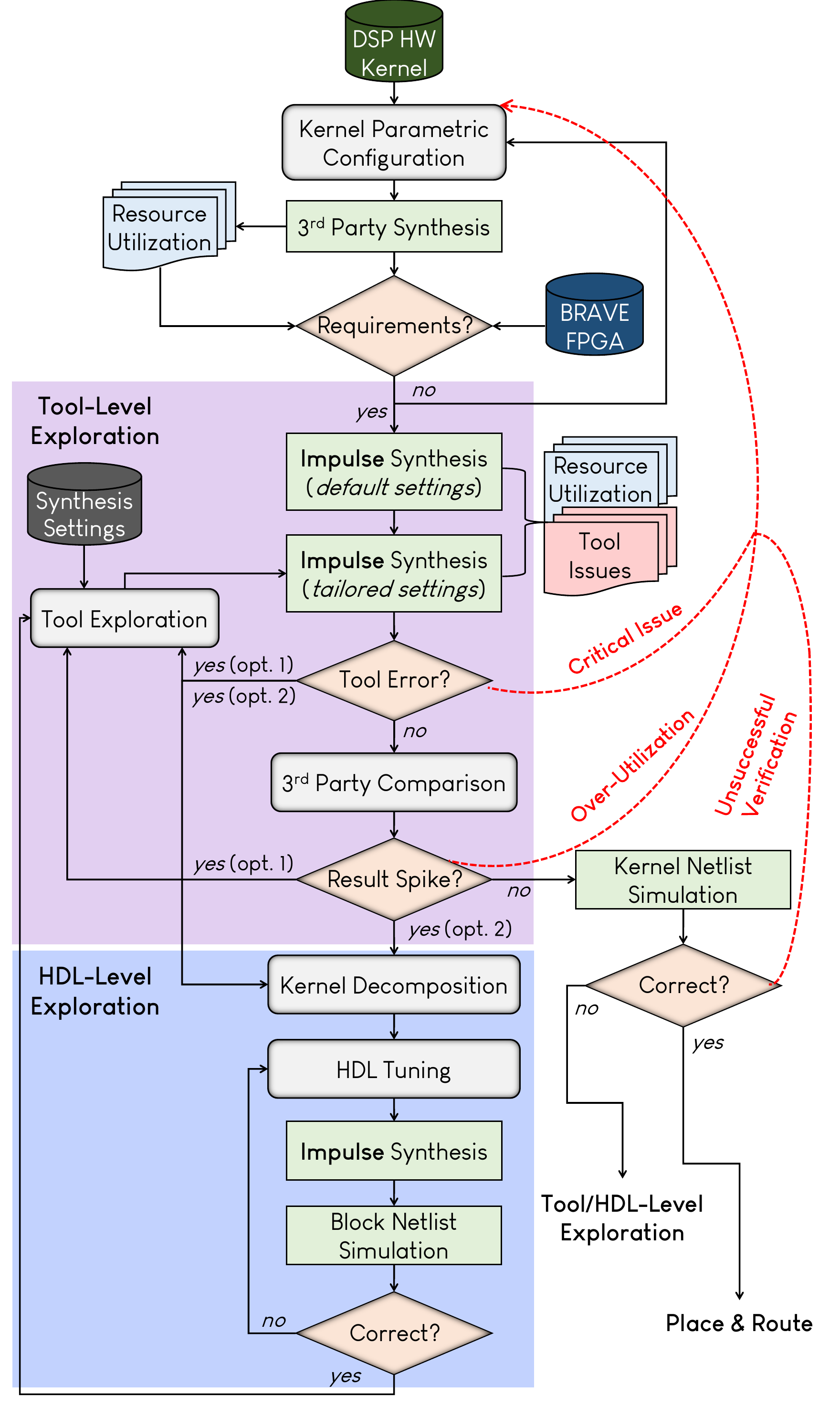}
\caption{Methodology for synthesizing DSP kernels on BRAVE FPGAs.}
\label{s_metho}
\end{figure}

The methodology for synthesis is illustrated in Fig. \ref{s_metho}
and consists of three main phases: 
the parametric configuration of the DSP kernels,
the exploration of the SW tool's settings, and the HDL-level exploration of the DSP kernels. 
In the first phase, 
algorithmic parameters
such as the input size, 
input partition, convolution size, datapath bit-width, and parallelization factor, 
are tuned 
with respect to the
architecture and resources of the targeted BRAVE FPGA. 
Then, 
syntheses are executed on 3rd-party vendors' tools to obtain baseline  results. 

In the tool-level exploration phase,
a preliminary synthesis is executed using the default tool settings to identify potential issues.
The next step is to examine all the Impulse tool's synthesis settings,
apply them in both standalone and combinatorial fashion, 
and evaluate the generated 
synthesis netlists.
Indicatively, we test tool settings
related to  
synthesizer's mapping effort,
register duplication,
arithmetic/logic/memory mapping targets, 
DSP utilization ratio, 
and encoding style of FSMs. 
If a tool error or unexpected result is observed, 
we attempt to resolve it through  
alternative tool settings or HDL coding.
The resource utilization of the most efficient, error-free netlist
is compared to that of the 3rd-party FPGA vendors.
In case of significant differentiation in the results, 
we adjust the tool settings or proceed to the next methodology phase.

In the HDL-level exploration phase, 
we break down the kernel into smaller building blocks to test them individually in a recursive manner. 
In this way, 
we thoroughly examine optimization issues and/or errors that would otherwise be challenging to detect within the entire kernel.
In this lower-level exploration,
we use 
standard template-based coding
and 
NanoXplore's HDL templates
to implement the building blocks. 
Moreover, 
post-synthesis simulation is performed for each HDL block that is examined. 
Finally, 
when all the error and/or optimization issues
are resolved,
we return to the tool-level exploration to evaluate again the entire kernel. 
In case the tool-level and the HDL-level explorations cannot deliver an acceptable design solution
due to tool error, resource over-utilization, or unsuccessful verification/simulation,
we use our methodology's 
feedback loops (red dashed lines in Fig. \ref{s_metho}) 
to re-execute all the phases from the start 
(i.e., with a new kernel configuration).

\begin{figure}[t]
\centering
\includegraphics[width=\columnwidth]{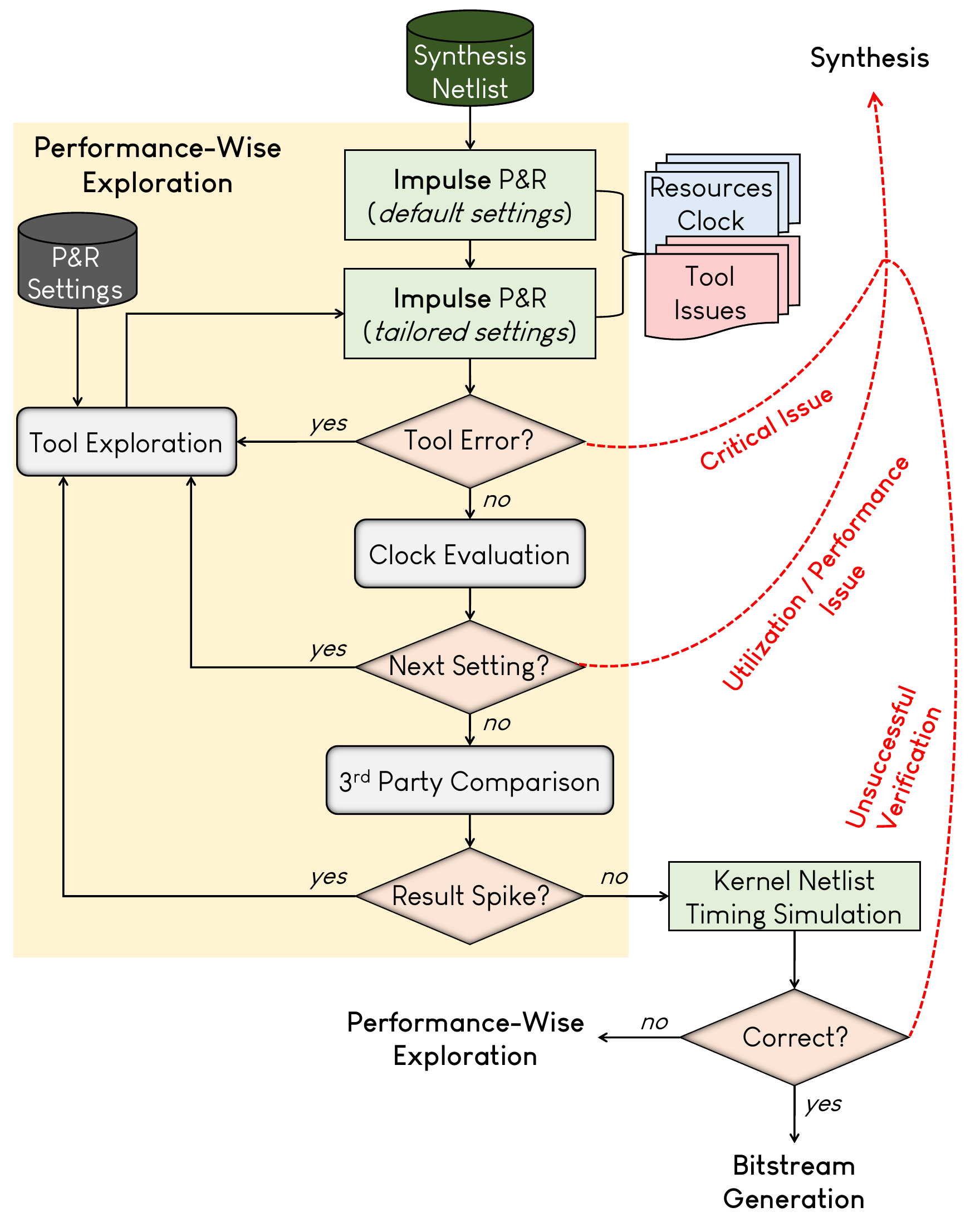}
\vspace{-15pt}
\caption{Methodology for placing \& routing DSP kernels on BRAVE FPGAs.}
\label{p_metho}
\end{figure}

Fig. \ref{p_metho} illustrates the corresponding methodology for the 
place \& route (implementation) tool stage.
It inputs the error-free synthesis netlist
and performs a performance-wise exploration of the tool settings
to extract  
the best achievable clock frequency. 
More specifically, 
we employ settings involving 
placement/routing and physical constraints,
as well as we evaluate   
location-specific placements 
by defining mapping
regions on the FPGA floorplan.
Furthermore, 
we adjust the timing-related settings
(e.g., constraints for clock, false path, and max delay, timing driven placement \& routing). 
Similarly to synthesis, we perform comparisons to 3rd-party tools, simulations, and use feedback loops for critical issues. 

\subsection{SoC's Processing System}

All the main features of the DAHLIA SoC, which is embedded in the NG-ULTRA devices, are targeted for evaluation. 
DAHLIA represents the first time NanoXplore integrates a SW-based SoC into their radiation-hardened FPGAs. 
Thereby, the maturity of the tools for the SW development is lower than those used for the RTL development. 
NanoXplore offers a SW Development Kit (SDK) that provides the required functionalities for executing bare-metal applications on DAHLIA in an efficient manner.

The proposed methodology is based on the Dhrystone and CoreMark synthetic benchmarks, 
which have been extensively used by the industry.
CoreMark results are preferred over Dhrystone as they are better formalized, while CoreMark was also developed more recently, and thus it incorporates more representative workloads.
For execution on the ARM Cortex-R52 processor,
we consider two set of compilation flags:

\begin{itemize}
    \item Set 1, no-inline: -O2, --no-inline.
    \item Set 2, inlining: no-inline: -O2, --no-inline. Set 2, inlining: -O2, funroll-loops, -funroll-all-loops, -finline, -finline-functions, -finline-limit=800, -falign-functions=4, falign-jumps=4, falign-labels=4, falign-loops=4, -flto, -fwhole-program.
\end{itemize}

\section{Experimental Evaluation}

For the experimental evaluation, we employ several compute- and memory-intensive DSP kernels, which are all used in modern space applications.
More specifically, 
Harris Corner Detector and Canny Edge Detector are used for feature detection (e.g., in vision-based navigation and docking applications),
GAD-Disparity and SpaceSweep are used for stereo matching (depth extraction in 3D
scene reconstruction),
Relative Navigation is used for spacecraft's landing operations,
and QAM Modulator \& Demodulator is used in the 5G waveform of non-terrestrial networks (satellite communications).   
The computer vision kernels are implemented for 1-Megapixel input images, while the telecom kernel (QAM) performs 16-order modulation \& demodulation. 
For comparison purposes, we perform runs on 3rd-party FPGAs with similar resources (AMD/Xilinx's Zynq UltraScale+ ZU09CG, Microchip's PolarFire MPFS460T, Intel's Arria 10 SX480), as well as on the smaller NanoXplore's FPGAs (NG-MEDIUM, NG-LARGE). 
The SW tool for the European FPGAs is Impulse v23.5.1.2.

Table 1 reports the experimental results for NG-ULTRA, including average values for the 3rd-party FPGAs.
For NG-ULTRA, the reported LUT values are derived from the sum of the 4-LUT plus the carry logic units. 
For the average 3rd-party metric, we extrapolate the LUT utilization of each device to 4-input LUTs (UltraScale+ and Arria have 6-LUT and 8-LUT architectures, respectively). 
The extrapolation is performed based on statistics exported from the implementations:
one 6-LUT corresponds to $\sim$1.5 4-LUT and 
one 8-LUT corresponds to $\sim$2 4-LUT. 
The NG-ULTRA results are the optimized ones,
which are produced   
based on our systematic exploration.
It is also worth noting that the clock frequency of NG-ULTRA is lower than that of the 3rd-party,
as its FPGA fabric is radiation-hardened,
impacting the critical paths and therefore the max clock frequency. 

For Harris, NG-ULTRA is slightly above (+6.6\%) the 3rd-party average LUT metric. DFFs and DSPs are comparable, with NG-ULTRA being more efficient in DFFs. 
For the block memory (RAMB) utilization, 
NG-ULTRA uses significantly less (108 vs. 182), but this is attributed to the bigger size of NG-ULTRA's RAMBs.
We also note that a mapping directive to use exclusively DSPs for multipliers and carry logic for adders is employed to boost the clock frequency (roughly 8.7\% better clock is achieved compared to default mapping).

\begin{table*}[t]
\renewcommand{\arraystretch}{1.15}
\setlength{\tabcolsep}{7.3pt}
\caption{Resource Utilization of DSP Kernels on the Rad-Hard NG-ULTRA FPGA and 3rd-Party FPGAs (Average Values)}
\vspace{-3pt}
\label{tb_res}
\centering
\begin{threeparttable}
\begin{tabular}{cccccccccccc}
\hline 
&  & \multicolumn{2}{c}{\textbf{LUT}}  & \multicolumn{2}{c}{\textbf{DFF}} & \multicolumn{2}{c}{\textbf{DSP}} & \multicolumn{2}{c}{\textbf{RAMB}} & \multicolumn{2}{c}{\textbf{MHz{\tiny\tnote{*}}}}\\
\cmidrule(lr){3-4}\cmidrule(lr){5-6}\cmidrule(lr){7-8}\cmidrule(lr){9-10}\cmidrule(lr){11-12}
    \multicolumn{1}{c}{\textbf{Kernel}}   &  \multicolumn{1}{c}{\textbf{Input Partition, I/O bits}}  & \emph{NG-U} & \emph{3rd-P} & \emph{NG-U} & \emph{3rd-P}  & \emph{NG-U} & \emph{3rd-P}  & \emph{NG-U} & \emph{3rd-P}  & \emph{NG-U} & \emph{3rd-P} \\

\hline \hline 
Harris          & 1024$\times$32,   8/32b & 16882 & 15831 & 19470 & 21016 & 92 & 81 & 108 & 182 & 80 & 183 \\
Canny           & 1024$\times$1024, 8/4b  & 2090 & 5951 & 2029 & 5631 & 0 & 0 & 177 & 293 & 67 & 170 \\
SpaceSweep      & 1024$\times$32,   8/32b & 11882 & 17333 & 10788 & 15905 & 50 & 36 & 278 & 401 & 58 & 188 \\
GAD-Disparity   & 1024$\times$32,   8/10b & 1306 & 12817 & 3838 & 14369 & 494 & 165 & 105 & 169 & 70 & 217 \\
RelNav          & 1024$\times$32, 8/--b{\tiny\tnote{**}} & 114489 & 96584 & 55083 & 56367 & 78 & 84 & 240 & 296 & 19 & 100  \\
QAM16           & $N$$\times$8, 8/8b & 19818 & 25119 & 29193 & 22463 & 256 & 162 & 0 & 0 & 92 & 303 \\
\hline
\end{tabular}
\begin{tablenotes}[flushleft]
   \item[\phantom{*}*]{\fontsize{6.3}{7.7}\selectfont Clock frequency of NG-ULTRA is significantly lower, as it is a radiation-hardened SoC FPGA (contrary to the 3rd-party FPGAs).}
    \item[**]{\fontsize{6.3}{7.7}\selectfont RelNav output is composed of a list of 100 image features (1600 bytes) per input image.}
   \end{tablenotes}
 \end{threeparttable}
 \vspace{-10pt}
\end{table*}

\begin{table}[!t]
\renewcommand{\arraystretch}{1.15}
\setlength{\tabcolsep}{8pt}
\caption{Throughput* of DSP Kernels on the EU Rad-Hard FPGAs}
\vspace{-3pt}
\label{tb_2}
\centering
\begin{threeparttable}
\begin{tabular}{c|ccc}
\hline 
\multicolumn{1}{c|}{\textbf{Kernel}}  & \textbf{NG-MEDIUM}  & \textbf{NG-LARGE} & \textbf{NG-ULTRA}  \\
\hline \hline 
Harris           & --{\tiny\tnote{**}}   & 4 FPS   & 11 FPS   \\
Canny            & 15 FPS   & 9 FPS   & 21 FPS   \\
SpaceSweep       & 92 MPDS  & 67 MPDS & 129 MPDS \\
GAD-Disparity    & --{\tiny\tnote{**}} & 7 MPDS & 20 MPDS \\
RelNav          & --{\tiny\tnote{**}}   & --{\tiny\tnote{**}}   & 0.8 FPS   \\
QAM16            & 454 MBPS  & 1415 MBPS  & 2959 MBPS  \\
\hline 
\textbf{Avg. Clock}           & 49 MHz  & 30 MHz  & 65 MHz  \\
\hline
\end{tabular}
\begin{tablenotes}[flushleft]
   \item[\phantom{*}*]{\fontsize{6.3}{7.7}\selectfont FPS: Frames Per Sec, MPDS: Megapixel Disparities Per Sec, MBPS: Megabits Per Sec.}
    \item[**]{\fontsize{6.3}{7.7}\selectfont Resource over-utilization for the targeted accuracy (even for smaller input image).}
   \end{tablenotes}
 \end{threeparttable}
 \vspace{-4pt}
\end{table}

For Canny, all the FPGAs do not utilize DSPs.
NG-ULTRA uses around 3$\times$ and 2.8$\times$ less LUTs and DFFs, respectively. 
This is mainly attributed 
to the additional logic utilized in PolarFire,
which increases the averaging 3rd-party value. 
The RAMB utilization of NG-ULTRA is also better (at $\sim$60\% of the average 3rd-party value).
The reported clock frequency in NG-ULTRA is achieved with manual pad placement in an effort to boost it even further (+2.6\% higher clock is achieved). 


For SpaceSweep, more or less the same picture is displayed: 
lower utilization in LUTs, DFFs, and RAMBs, but with the outlier of DSP utilization 
($\sim$38\% higher than the 3rd-party one).
For GAD-Disparity, we see a massive decrease in NG-ULTRA's LUTs and DFFs (1306 vs. 12817 and 3838 vs. 14369, respectively). 
This is easily explained by our design choice to map all the multipliers and adders to DSPs (494 vs. 165) in order to the boost clock frequency.
On the other hand, for RelNav, NG-ULTRA is worse in LUTs but better in DSPs. 

Finally, for QAM16, we observe lower LUT utilization and higher DFF and DSP utilization in NG-ULTRA. 
By employing manual pad placement, an improvement of +6\% is achieved in clock frequency. Although QAM16 is the best performer for NG-ULTRA in terms of raw clock frequency ($\sim$92MHz), it is 
still around 1/3 of the 3rd-party's frequency,
which is a ratio that 
is followed in almost all kernels. 

Table \ref{tb_2} reports the throughput results of 
our DSP kernels on the NanoXplore's rad-hard FPGA variants. 
The kernel configuration is as reported in Table \ref{tb_res}, except for NG-MEDIUM,
where the 1-MegaPixel input image is
divided into 
4 512$\times$512 images for Canny
and 16 256$\times$256 images for SpaceSweep,   
and they are executed serially. 
For NG-MEDIUM, the parallelization factor of QAM16 is also reduced from 8 to 2.  
For NG-LARGE, SpaceSweep processes successively 4 512$\times$512 input images.
Moreover, it is important to note that
some kernels do not fit in NG-MEDIUM and NG-LARGE for the targeted internal and output accuracy, even for smaller input images (LUTs/DSPs/CYs are over-utilized). 

\begin{table}[!t]
 \vspace{-1.4pt}
\renewcommand{\arraystretch}{1.15}
\setlength{\tabcolsep}{5.8pt}
\caption{Benchmarking of NG-ULTRA's Processing System (Cortex-R52)}
\vspace{-3pt}
\label{tb_3}
\centering
\begin{threeparttable}
\begin{tabular}{c|cc}
\hline 
\multicolumn{1}{c|}{\textbf{Metric}}  & \textbf{Compilation Flags Set 1}  & \textbf{Compilation Flags Set 2} \\
\hline \hline 
DPS{\tiny\tnote{*}}             &  1496259 & 2643171    \\
DMIPS{\tiny\tnote{*}}           &  851.6 & 1504.37     \\
DMIPS/MHz                       &  1.42 & 2.51     \\
\hline
\end{tabular}
\begin{tablenotes}[flushleft]
    \item[*]{\fontsize{6.3}{7.7}\selectfont DPS: Dhrystones Per Sec, DMIPS: Dhrystone Millions of Instructions Per Sec.}
   \end{tablenotes}
 \end{threeparttable}
 \vspace{-4pt}
\end{table}

For Canny and SpaceSweep, the throughput advantage of NG-MEDIUM over NG-LARGE is attributed solely to the clock frequency difference, as NG-MEDIUM completes faster the equivalent runs of smaller images compared to a full-sized one in NG-LARGE. 
NG-ULTRA achieves around 2.5--3$\times$ the throughput of NG-LARGE. 
For Canny and SpaceSweep, NG-ULTRA provides an 40\% increase in throughput compared to the serially extrapolated runs of NG-MEDIUM. 
These throughput results correlate with the average clock frequency, which is 1.4$\times$ higher in NG-ULTRA compared to NG-MEDIUM, and nearly 2.2$\times$ higher compared to NG-LARGE.
Overall, the kernel's throughput is considered sufficient considering the expected performance of relevant space applications \cite{aiaa}. 
Nevertheless, 
custom re-design of the kernels 
(e.g., with higher parallelization and manual placement),
tailored to the bigger NG-ULTRA,  
is expected to deliver even better throughput results. 

Finally, 
Table \ref{tb_3} reports the results for the 
Dhrystone benchmarking of the SoC's ARM processor.  
The respective benchmarking scores for CoreMark are up to 3.108 CoreMarks/MHz. 
These results fit within the expected margin for the evaluated processors, even when they are compared with non-radiation-hardened implementations \cite{eembc}. 

\section{Conclusion}
In this paper, 
we presented a systematic approach to design, test, and evaluate 
high-performance DSP algorithms on the new European radiation-hardened FPGAs. 
Based on our methodology,
we delivered competitive, resource-balanced implementations on NG-ULTRA with sufficient performance,
which has been significantly improved from the predecessor devices.  
Our future work will focus on the HW/SW co-design of space applications on the NG-ULTRA SoC FPGA.  

\bibliographystyle{IEEEtran}
\bibliography{REFERENCES.bib}

\begin{thebibliography}{10}
\providecommand{\url}[1]{#1}
\csname url@samestyle\endcsname
\providecommand{\newblock}{\relax}
\providecommand{\bibinfo}[2]{#2}
\providecommand{\BIBentrySTDinterwordspacing}{\spaceskip=0pt\relax}
\providecommand{\BIBentryALTinterwordstretchfactor}{4}
\providecommand{\BIBentryALTinterwordspacing}{\spaceskip=\fontdimen2\font plus
\BIBentryALTinterwordstretchfactor\fontdimen3\font minus \fontdimen4\font\relax}
\providecommand{\BIBforeignlanguage}[2]{{%
\expandafter\ifx\csname l@#1\endcsname\relax
\typeout{** WARNING: IEEEtran.bst: No hyphenation pattern has been}%
\typeout{** loaded for the language `#1'. Using the pattern for}%
\typeout{** the default language instead.}%
\else
\language=\csname l@#1\endcsname
\fi
#2}}
\providecommand{\BIBdecl}{\relax}
\BIBdecl

\bibitem{aiaa}
G.~Lentaris \emph{et~al.}, ``{High-Performance Embedded Computing in Space: Evaluation of Platforms for Vision-Based Navigation},'' \emph{AIAA Journal of Aerospace Information Systems}, vol.~15, no.~4, pp. 178--192, 2018.

\bibitem{nxx}
E.~Lepape and M.~Le~Penven, ``{New Generation of Rad-Hard SoC FPGA},'' in \emph{Space FPGA Users Workshop (SEFUW)}, 2023, pp. 1--28.

\bibitem{spacebrave}
R.~D. Torrijos, ``{GMV and BRAVE FPGAs: From Studies to Flight Hardware Use},'' in \emph{NanoXplore's BRAVE Days}, 2023, pp. 1--32.

\bibitem{ngmoon}
\BIBentryALTinterwordspacing
NanoXplore, \emph{NG-MEDIUM and FUSIO-RT: Mission completed on the dark side of the Moon}, (Accessed: 8-6-2024). [Online]. Available: \url{https://nanoxplore.com/index.php/2024/06/07/mediumfusioonmoon/}
\BIBentrySTDinterwordspacing

\bibitem{ultra}
N.~Ibellaatti \emph{et~al.}, ``{HERMES: Qualification of High Performance Programmable Microprocessor and Development of Software Ecosystem},'' in \emph{Design, Automation \& Test in Europe (DATE)}, 2023, pp. 1--5.

\bibitem{dahlia}
J.~Poupat, ``{DAHLIA: Very High Performance Microprocessor for Space Applications},'' in \emph{ESA Workshop on Avionics, Data, Control and Software Systems (ADCSS)}, 2017, pp. 1--32.

\bibitem{ahs_brave}
K.~Maragos \emph{et~al.}, ``{Evaluation Methodology and Reconfiguration Tests on the New European NG-MEDIUM FPGA},'' in \emph{NASA/ESA Conf. on Adaptive Hardware and Systems (AHS)}, 2018, pp. 127--134.

\bibitem{access_brave}
V.~Leon \emph{et~al.}, ``{Development and Testing on the European Space-Grade BRAVE FPGAs: Evaluation of NG-Large Using High-Performance DSP Benchmarks},'' \emph{IEEE Access}, vol.~9, pp. 131\,877--131\,892, 2021.

\bibitem{shyloc_fpga}
Y.~Barrios \emph{et~al.}, ``{SHyLoC 2.0: A Versatile Hardware Solution for On-Board Data and Hyperspectral Image Compression on Future Space Missions},'' \emph{IEEE Access}, vol.~8, pp. 54\,269--54\,287, 2020.

\bibitem{zenoleon}
V.~Leon \emph{et~al.}, ``{Systematic Evaluation of the European NG-LARGE FPGA \& EDA Tools for On-Board Processing},'' in \emph{European Workshop on On-Board Data Processing (OBDP)}, 2021, pp. 1--8.

\bibitem{klemen_new}
K.~Bravhar \emph{et~al.}, ``{SerDes Integrated Into the SpaceWire Interface Helps in Achieving Higher Data Rates},'' \emph{IEEE Aerospace and Electronic Systems Magazine}, vol.~38, no.~11, pp. 16--27, 2023.

\bibitem{cnn_brave}
A.~Portaluri \emph{et~al.}, ``{Design Techniques for Multi-Core Neural Network Accelerators on Radiation-Hardened FPGAs},'' in \emph{Int'l. Symposium on Parallel and Distributed Computing (ISPDC)}, 2023, pp. 16--22.

\bibitem{obc_ultra}
E.~Danard and A.~Comolet-Tirman, ``{OBC-Ultra, the Rad-Hard NG-Ultra-based On Board Computer for Future Applications},'' in \emph{European Data Handling \& Data Processing Conference (EDHPC)}, 2023, pp. 1--4.

\bibitem{eembc}
\BIBentryALTinterwordspacing
EEMBC, \emph{PIC32MM0064GPL036 Benchmark Score Viewer}, (Accessed: 7-6-2024). [Online]. Available: \url{https://www.eembc.org/viewer/?benchmark{\_}seq=2538}
\BIBentrySTDinterwordspacing

\end{thebibliography}

\end{document}